Diamond chemical vapor deposition on optical fibers for fluorescence waveguiding

J. R. Rabeau[a)], S. T. Huntington, A. D. Greentree, and S. Prawer

*School of Physics, University of Melbourne, Victoria, Australia, 3010*

a) Corresponding author email: jrabeau@unimelb.edu.au

Abstract

A technique has been developed for depositing diamond crystals on the endfaces of optical fibers and capturing the fluorescence generated by optically active defects in the diamond into the fiber. This letter details the diamond growth on optical fibers and transmission of fluorescence through the fiber from the nitrogen-vacancy (N-V) color center in diamond. Control of the concentration of defects incorporated during the chemical vapor deposition (CVD) growth process is also demonstrated. These are the first critical steps in developing a fiber coupled single photon source based on optically active defect centers in diamond.

The availability of precise silica fabrication technologies, flexible waveguiding technologies and high speed switching have led to many successes in photonics research and industry. In further building on such successes, the next major breakthrough in photonics is set to occur with the integration of true quantum elements with classical photonic structures. As such, these devices would build upon, but are quantitatively different from classical optically active structures such as difference frequency generation (DFG) lasers. Towards this aim, we report here the incorporation of optically active centers in diamond within waveguiding structures. In particular we report the growth of thin diamond films and microcrystals containing the [N-V]$^-$ center on endfaces of optical fibers and demonstrate that the fluorescence emission from the optically active centers can be coupled directly into the fiber.

Growth on the optical fiber brings the optically active centers into the near-field for coupling to the fiber and opens the possibility of devices where the centers are used as coupling elements (for example on an NSOM probe for surface analysis), or where the fluorescence is the desired product, for example as a single photon source. As such, these developments should be seen as being entirely complimentary to the recent work employing N-V diamond as a single photon source[1] and developments in quantum dot single photon sources.[2,3,4]

N-V centers form readily in the high temperature regimes (600 to 1200°C) of the chemical vapor deposition (CVD) of diamond. It is highly stable, optically active (at room temperature) and the spectral properties of these centers have been well studied.[5] The N-V center was therefore selected to demonstrate the fabrication of an integrated diamond photonic device using CVD. Growing diamond on the fiber endface effectively positions the optically active defect centers in the waveguide mode field which enhances the coupling and eliminates cumbersome optical components and alignment problems. Diamond growth on optical fibers was previously investigated[6] for wear and chemical resistance applications, and N-V containing diamond nanoparticles have been glued to optical probe tips and used as nanoscopic light sources by externally exciting the particles with a laser.[7] The present work employs CVD diamond growth on optical fibers for the purpose of fluorescence waveguiding.

CVD diamond was grown with varying concentrations of N-V color centers by adjusting the nitrogen doping levels in a standard CVD process. The CVD reactor consisted of a standard bell-jar growth chamber with a 1.5 kW microwave power supply (ASTeX). FIG. 1 shows a series of photoluminescence (PL) spectra of diamond grown on silicon substrates in the CVD reactor. The diamond films were grown with 800°C substrate temperature, 1.5 kW microwave power, 50 Torr and 0 – 0.1% $N_2$, 0.7% $CH_4$, ~99.2% $H_2$ gas mixtures for ~20 hours. Under 514 nm laser excitation, a clear rise in the [N-V]⁻ photoluminescence was observed as a function of the nitrogen composition of the gas mixture. This demonstrates the ability to control the concentration of optically active centers in the diamond and therefore the fluorescence intensity. Single defects can be fabricated by optimizing the nitrogen doping levels.

Diamond crystals were then grown on optical fiber endfaces using the CVD reactor. Fiber substrates were prepared by cleaving (Furukawa fiber cleaver) and bundling 50 to 100, 1.5 cm long fibers and seeding by exposing the fibers to ultrasonification[8] in a diamond/metal powder slurry. The samples were then loaded into the reactor chamber, with the scratched end of the cleaved fiber bundle facing up towards the plasma. CVD conditions for diamond growth on fibers were typically 1.2 kW microwave power, 30 Torr total pressure and ~99.2% $H_2$, 0.7 % $CH_4$ and <0.1% $N_2$ composition. The substrate temperature was maintained at 700°C for the duration of the 4 hour growth runs.

Scanning electron microscopy (SEM) was performed on the processed fiber bundles after coating with ~20 nm of gold. FIG. 2 shows a series of images including a bundle of optical fibers, a single fiber and the core region of a single fiber.

The presence of the N-V centers on the fiber endfaces was confirmed using optical spectroscopy. Raman and PL spectra of the diamond grown on the fiber endfaces were collected using a Raman/PL spectrometer (Renishaw 2000) with 514 nm excitation ($Ar^+$ Spectra Physics).

Of considerable interest were optical transmission experiments that verified the ability to couple and guide the fluorescence from diamond crystals through the optical fiber. Fibers with diamond crystals located on the fiber core were selected via optical microscope inspection and the non-diamond-coated end was fusion spliced (Ericsson FSU 995) with a clean optical fiber several tens of centimeters in length. Using a 100 X objective optical fiber coupler, a 10 mW 514 nm laser beam was focused onto the core-diamond by aligning the fiber coupler for maximum throughput of the excitation beam (estimated by observing the visible throughput of the pump beam). The opposing end of the fiber was then directed into the spectrometer via a 100 X light collection objective lens.

FIG. 3 shows a room temperature PL spectrum collected by reflection from the CVD diamond on the optical fiber endface compared to a PL spectrum transmitted through the optical fiber. The relevant peaks are labeled showing the characteristic Raman features expected from CVD grown diamond including the $sp^3$ and $sp^2$ carbon peaks at 552 and ~ 558 nm respectively. The [N-V]$^-$ zero phonon line (ZPL) occurs at 637 nm, with phonon sidebands extending to ~ 720 nm. A normalized PL spectrum from a bare optical fiber excited identically to the diamond coated fibers is also included for comparison. The peaks at around 550 and 580 nm in the bare fiber spectrum are most likely fluorescence features emanating from the fused silica of the optical fiber and are not due to the diamond. The optical fiber employed in these experiments (ThorLabs, FS-SN-3224, NA 0.12) had a cutoff frequency of 620 ± 50 nm and is designed for single-moded operation at 630 nm. Oscillations between 650 and 750 nm in the transmitted spectrum are believed to be interference patterns generated from the diamond thin film. It is noted that the general shape of the spectra and the intensity ratios of the $sp^3$ and $sp^2$ carbon and the N-V fluorescence are faithfully reproduced after coupling into and transmitting along the fibre. . This gives strong evidence for the fluorescence waveguiding.

In summary, the controlled growth of diamond on optical fibers has been demonstrated indicating strong potential for the fabrication of a range of efficient and robust integrated photonic devices. By adjusting the doping parameters in the CVD growth process, control over the density of optically active centers in diamond was achieved. Diamond was grown on optical fiber endfaces, and the diamond Raman features as well as photoluminescence from $[N-V]^0$ (575 nm) and $[N-V]^-$ (637 nm) were observed. Photoluminescence was detected from both ends of the fiber, i.e. directly from the fiber endface and propagated through the optical fiber. In this experiment, waveguiding of the N-V center fluorescence was shown, however a wide range of optically active centers in diamond are known[9], the only limitation being the ability to incorporate these centers in the CVD process. This is an important step towards the fabrication of an optical fiber integrated diamond single photon source.

FIG. 1.  PL spectra of diamond grown with an increasing amount of nitrogen in the process gas mixture.  The concentration of [N-V]$^-$ centers increases as a function of nitrogen doping.  These spectra were collected at liquid nitrogen temperature, and therefore show narrowing of the peak widths, a result of a restriction in vibrational relaxation pathways.

FIG. 2.  SEM images of diamond grown on optical fiber endfaces.  The bottom image shows diamond grains located on the core of the optical fiber.

FIG. 3. Direct collection and transmitted collection of photoluminescence from CVD diamond grown on an optical fiber.  The PL spectrum from a bare fiber is shown for comparison.  The bare fiber spectrum and diamond coated fiber spectrum were normalized using the intensity of the side band of the excitation laser.  The direct and transmitted spectra were approximately normalized using the intensity of the diamond Raman transition at 552 nm.  a) shows the full photoluminescence spectrum and b) shows a magnification of the diamond and graphite Raman components.

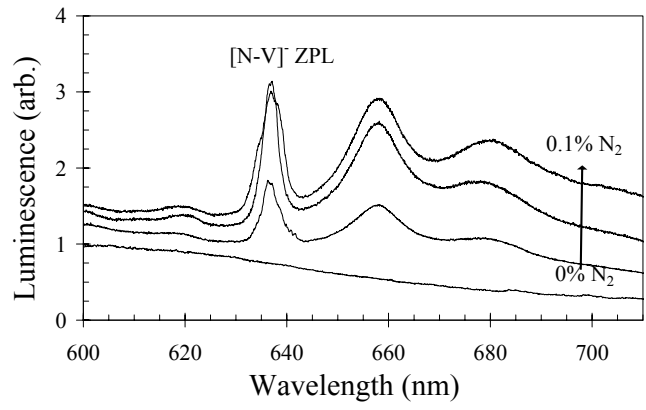

FIG. 1. J. R. Rabeau

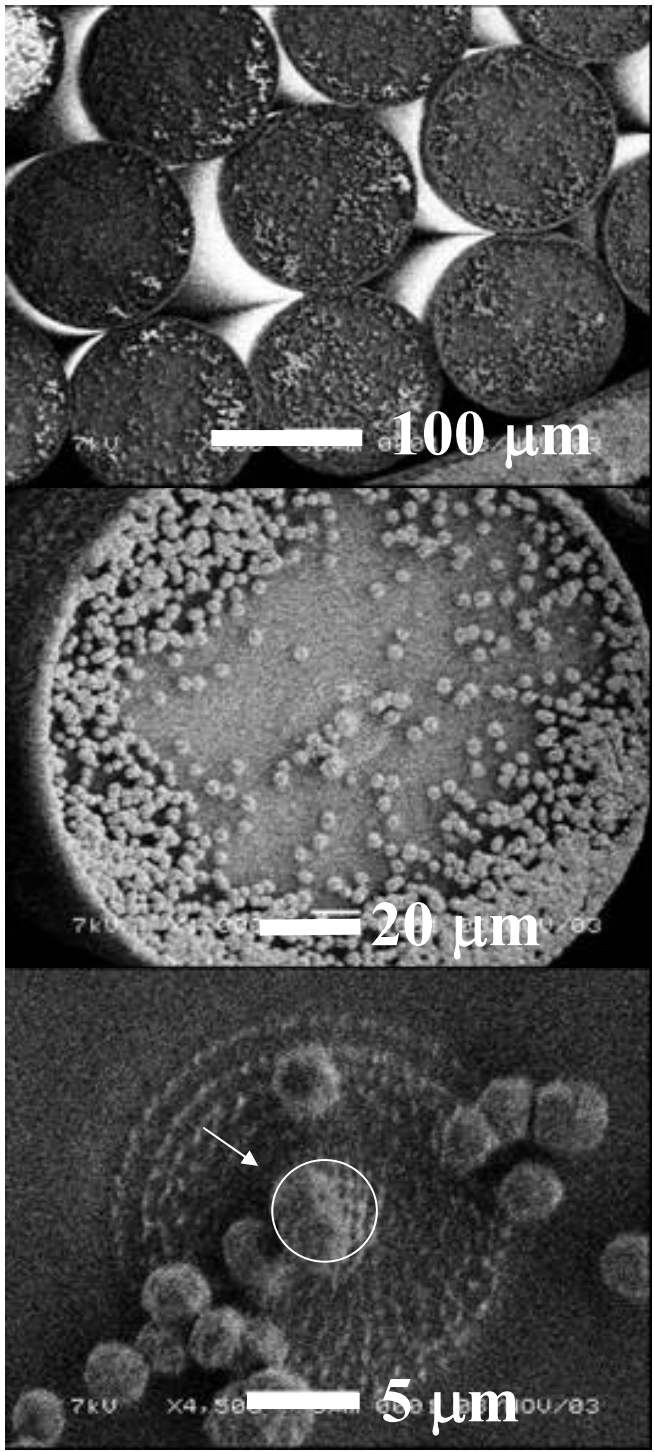

FIG. 2. J. R. Rabeau

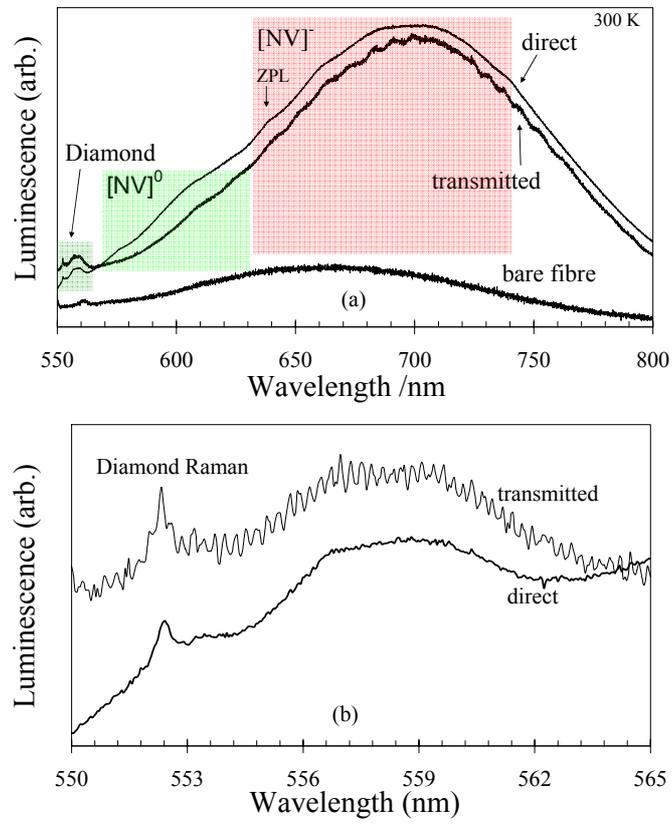

FIG. 3. J. R. Rabeau